\documentclass[aps,pra,amssymb,amsmath,eqsecnum,nofootinbib]{revtex4}
\usepackage{graphicx}

\newtheorem{definition}{Definition}[section]

\newtheorem{proposition}{Proposition}[section]

\newtheorem{remark}{Remark}[section]
\newtheorem{axiom}{AXIOM}[section]

\newenvironment{hypothesis}{HP: \begin{center}} {\end{center}}
\newenvironment{thesis}{TH: \begin{center}} {\end{center}}
\newenvironment{proof}{\begin{center}PROOF: \end{center}} {$ \blacksquare $}

\begin{document}
\title{Experimental indications of the  violation of the Martin L\"{o}f - Solovay - Chaitin  Axiom
of Reduction of Quantum Mechanics and of  the necessity of
replacing it with the weaker Von Mises -  Church Axiom of
Reduction}
\author{Gavriel Segre}
\begin{abstract}
Under the assumption that the quantum  random number generator Quantis by
Id Quantique is a fair quantum coin, experimental indications of
the fact that independent tosses of a quantum coin don't pass the
Iterated Logarithm Randomness' Test are shown.

The consequential observed experimental violation of the Martin
L\"{o}f - Solovay - Chaitin Axiom of Reduction of Quantum
Mechanics is then interpreted as the indication of  the necessity
of replacing it with the weaker Von Mises - Church Axiom of
Reduction whose Game Theoretic meaning is explained.
\end{abstract}
\maketitle
\newpage
\tableofcontents
\newpage
\section{Acknowledgements}
I would like first of all to thank Cristian S. Calude for many precious remarks.

I would then like to thank strongly Vittorio de Alfaro for his
friendship and his moral support, without which I would have
already given up.

Finally I  would like to thank strongly Andrei Khrennikov and the
whole team at the International Center of Mathematical Modelling
in Physics and Cognitive Sciences of V\"{a}xj\"{o} for their very
generous informatics' support.

Of course nobody among the mentioned people has responsibilities
as to any (eventual) error contained in these pages.

\newpage
\section{The quantum coin, the Axiom of Reduction of Quantum Mechanics and the Law of the Iterated
Logarithm} \label{sec:The quantum coin, the Axiom of Reduction of
Quantum Mechanics and the Law of the Iterated Logarithm}

Given a 1-qubit quantum system prepared in the state:
\begin{equation} \label{eq:unbiased state}
    | unbiased > \; := \; \frac{1}{ \sqrt{2} } ( | 0 > + | 1 > )
\end{equation}
let us introduce the \emph{qubit operator}:
\begin{equation}
    \hat{q} \; := \; - | 0 > < 0 | \, + \, | 1 > < 1 |
\end{equation}

Following the literature on the subject, I will refer to such a
quantum physical system as to the \emph{quantum coin}.

Let us call X the classical random variable defined as the result
of a measurement of the qubit operator when the system is  in the
state $ | unbiased > $.

Let us then  recall the following (see for instance
\cite{Sakurai-94}, \cite{Ballentine-98}):
\begin{axiom} \label{ax:Axiom of Reduction of Quantum Mechanics}
\end{axiom}
\emph{Axiom of Reduction of Quantum Mechanics (for a quantum coin)
}
\begin{enumerate}
    \item X is a
Bernoulli($\frac{1}{2}$) random variable such that:
\begin{equation}
    Pr[ X=- 1 ] \; = \; Pr[ X = 1 ] = \frac{1}{2}
\end{equation}
\begin{equation}
  Pr [ X \notin \{ 1 , - 1 \} ] \; = \; 0
\end{equation}
    \item if the result of the measurement of $  \hat{q} $ is
    equal to -1 then the state of the quantum coin after the
    measurement is $ |0 > $; if, contrary, the result of the measurement of $  \hat{q} $ is
    equal to 1 then the state of the quantum coin after the
    measurement is $ |1 > $.
\end{enumerate}

 A trivial computation shows that X has zero mean and unit
variance:
\begin{equation}
    E( X ) \; = \; ( -1 ) \frac{1}{2} + 1 \frac{1}{2} = 0
\end{equation}
\begin{equation}
    Var( X) \; := \; E [ ( X -  E( X ) )^{2} ] \; = \; ( -1)^{2}
    \frac{1}{2} + 1^{2} \frac{1}{2} \; = \; 1
\end{equation}

Let us now consider $ n \in \mathbb{N}_{+} $ copies of our quantum
system and let  $ X_{n} $ be the classical random variable defined
as the result of a measurement of the qubit operator on the $
n^{th} $ copy when such a system is in the state $ | unbiased > $
.

Clearly $ \{ X_{n} \}_{n \in \mathbb{N}_{+}} $ is a collection of
independent, identically distributed classical random variables
all having the same probability distribution of X.

Let us furthermore denote with $ x_{n} $ the experimental value of
the random variable $ X_{n} $.

The Law of Large Numbers states that:
\begin{proposition} \label{prop:law of large numbers}
\end{proposition}
\emph{Law of Large Numbers:}
\begin{equation}
    \lim_{n \rightarrow \infty} \frac{ \sum_{i=1}^{n} x_{i} }{n}
    \; = \; E(X) \; = \; 0
\end{equation}

The speed of such a convergence is ruled by the following
\cite{Kolmogorov-92}, \cite{Kolmogorov-56}, \cite{Billingsley-95},
\cite{Gut-05}:
\begin{proposition} \label{prop:law of the iterated logarithm}
\end{proposition}
\emph{Law of the Iterated Logarithm:}
\begin{equation}
    | \{ n \in \mathbb{N}_{+} \, : \,  \frac{\sum_{i=1}^{n} x_{i} }{\sqrt{2 n \log \log
    n}} \in [ 1 - \epsilon , 1 + \epsilon ] \} | \; = \; \aleph_{0} \;
    \; \;  \forall \epsilon >0
\end{equation}
where I have denoted by $ |S| $ the cardinality of a set S and
where I recall that a set has cardinality $ \aleph_{0} $ if and
only if it is infinite and countable.

\smallskip

As I will show in this paper, under the hypothesis that that the
quantum random number generator \emph{Quantis} realized by the
spin-off company of the University of Geneva \emph{id Quantique}
\cite{idQuantique-08} (based on the simplest beam splitter, i.e. a
semi-trasparent mirror \cite{Leonhardt-97}; for the credentials of
\emph{Quantis} see
\cite{Jennewein-Achleitner-Weichs-Weinfurter-Zeilinger-00},
\cite{randomnumbers-07} and \cite{Calude-05}) is a  fair quantum
coin, the statistical analysis of the experimental data concerning
repeated independent measurements of the qubit operator on a
one-qubit system prepared in the unbiased state shows that the
proposition \ref{prop:law of the iterated logarithm} is
experimentally violated.

 This implies the experimental violation of the
Reduction Axiom under which it has been deduced.

The natural framework in which to analyze the issue is Algorithmic
Information Theory \cite{Calude-02}.

In such a framework the first point of the Reduction Axiom of
Quantum Mechanics, i.e of the axiom \ref{ax:Axiom of Reduction of
Quantum Mechanics}, may be stated as the condition that the binary
sequence of the experimental results of measurements of the qubit
operator prepared in the unbiased state is Martin L\"{o}f -
Solovay - Chaitin random with probability one.

The mentioned experimental violation of such an axiom indicates
the necessity of  replacing it with a weaker axiom, the Von Mises
- Church Axiom of Reduction, stating that the binary sequence of
the experimental results of measurements of the qubit operator
prepared in the unbiased state is Von Mises  - Church random with
probability one.

I would like to stress that this fact doesn't contrast all the
extraordinary experimental confirmations that Quantum Mechanics
has received for more than a century:

these results don't involve the difference between the  Martin
L\"{o}f - Solovay - Chaitin Axiom of Reduction and the Von Mises -
Church Axiom of Reduction and can then been seen as confirmations
of both.

If, anyway, one is so  meticulous to choose an ad hoc experiment
distinguishing them, one observes the violation of the Martin
L\"{o}f - Solovay - Chaitin Axiom of Reduction and the necessity
of replacing it with the weaker Von Mises - Church Axiom of
Reduction.

\newpage
\section{Martin L\"{o}f - Solovay - Chaitin randomness}

Let us start from the following:
\begin{definition}
\end{definition}
\emph{alphabet:}
\begin{equation}
    \text{ a set A such that }  2 \leq |A| < \aleph_{0}
\end{equation}

Given an alphabet A let us denote by $ A^{+} := \cup_{n \in
\mathbb{N}_{+}} A^{n} $ the infinite countable set of all the
strings over A.

Let us then introduce on such a set the following useful partial
ordering:
\begin{definition}
\end{definition}
\emph{$ \vec{x} $ is a prefix of $ \vec{y} $:}
\begin{equation}
  \vec{x} <_{p}  \vec{y} \; := \; \exists \vec{z} \in A^{+} \;
  : \; \vec{y} = \vec{x} \cdot \vec{z}
\end{equation}

Given a set $ S \subset A^{+} $:
\begin{definition}
\end{definition}
\emph{S is prefix-free:}
\begin{equation}
 \vec{x} \nless_{p} \vec{y} \; \; \forall \vec{x} , \vec{y} \in S
\end{equation}

Let us now briefly recall some necessary rudiments of
Computability Theory \cite{Cutland-80}, \cite{Rogers-87},
\cite{Davis-Sigal-Weyuker-94}, \cite{Lewis-Papadimitriou-98}.

Given a partial function $ C : A^{+} \stackrel{\circ}{\mapsto}
A^{+} $ ,  we can look at it as a total function adding a
"non-halting" symbol (that following \cite{Calude-02} I will
assume to be the infinity symbol $ \infty $) and posing $ C (
\vec{x} ) := \infty $ whether C doesn't halt on $ \vec{x} $.

We can then introduce the following:
\begin{definition}
\end{definition}
\emph{halting set of C:}
\begin{equation}
    HALT_{C} \; := \; \{ \vec{x} \in A^{+} : C(\vec{x} ) \neq
    \infty \}
\end{equation}

We will say that:
\begin{definition}
\end{definition}
\emph{C is a partial recursive function:}
\begin{equation}
 \exists f :  \mathbb{N} \stackrel{\circ}{\mapsto}  \mathbb{N} \;
 \text{ partial recursive } : C(\vec{x}) =
 string(f(string^{-1}(\vec{x}))
\end{equation}
where $ string : \mathbb{N} \mapsto A^{+} $ is the map associating
to an integer n the $ n^{th} $ string of $ A^{+} $ in
lexicographic order and  where I demand to the mentioned
literature as to the definition of a partial recursive function on
natural numbers.

Given a set $ X \subset A^{+} $:
\begin{definition}
\end{definition}
\emph{X is computable:}
\begin{center}
 the characteristic function $ \chi_{X} $ of X is recursive
\end{center}
A notion weaker than computability is the following:
\begin{definition}
\end{definition}
\emph{X is computably enumerable:}
\begin{equation}
    \exists C : A^{+}  \stackrel{\circ}{\mapsto} A^{+} \text{ partial recursive
    function} \; : \; X = HALT_{C}
\end{equation}

We can at last introduce the following:

\begin{definition}
\end{definition}
\emph{Chaitin computer:}

\begin{center}
a partial recursive function $ C : A^{+} \stackrel{\circ}{\mapsto}
A^{+} $ such that $ HALT_{C} $  is prefix free.
\end{center}

\begin{definition}
\end{definition}
\emph{Universal Chaitin computer:}

a Chaitin computer U such that for every Chaitin computer C there
exists a constant $ c_{U,C} \in \mathbb{R}_{+} $ such that:
\begin{equation}
 \forall \vec{x} \in HALT_{C} , \exists \vec{y} \in  A^{+} \; :
 \; U ( \vec{y} ) = C ( \vec{x} ) \, \wedge \, | \vec{y} | \leq |
 \vec{x}| + c_{U,C}
\end{equation}
where $ \wedge $ is the logical conjunction (i.e. the logical
\emph{and operator}) and where $ |\vec{x}| $ is the length the
string $ \vec{x} $.

Given a universal Chaitin computer U and a string $ \vec{x} \in
A^{+} $ :
\begin{definition} \label{def:algorithmic information}
\end{definition}
\emph{algorithmic information of $ \vec{x} $ with respect to U:}
\begin{equation}
    I_{U} ( \vec{x} ) \; := \; \left\{%
\begin{array}{ll}
    \min \{ | \vec{y} | :  \vec{y} \in A^{+} , U( \vec{y} ) = \vec{x} \}, & \hbox{if $ \{ \vec{y} \in A^{+} : U( \vec{y} ) = \vec{x} \} \neq \emptyset $;} \\
    + \infty , & \hbox{otherwise.} \\
\end{array}%
\right.
\end{equation}

Let us now introduce the concept of \emph{algorithmic randomness}
(introduced in independent but equivalent ways by Martin-L\"{o}f,
by Solovay and by Chaitin) by defining the algorithmically random
sequences as those sequences with maximal algorithmic information
(and that hence cannot be algorithmically compressed):

\begin{definition} \label{def:Martin-Lof Solovay Chaitin  randomness}
\end{definition}
\emph{Martin L\"{o}f-Solovay Chaitin random sequences on A:}
\begin{equation}
    RANDOM_{MLSC} ( A^{\mathbb{N}_{+}} ) \; := \; \{ \bar{x} \in
 A^{\mathbb{N}_{+}} \, : \, \forall U \, \text{ universal Chaitin
    computer} \, \exists c_{U} \in ( 0 , + \infty ) \; :
    \;  I ( \vec{x}(n) ) \geq n - c_{U} \, \, \forall n \in
    \mathbb{N}_{+} \}
\end{equation}
where, given two sets A and B, $ A^{B} := \{ f : B \mapsto A \} $
is the sets of all the maps with domain B and codomain A, and
where $ \vec{x}(n) $ is the prefix of length n of the sequence $
\bar{x}$.

It may be proved that:

\begin{proposition} \label{prop:Martin Lof-Solovay Chaitin random sequences obey the
Law of the Iterated Logarithm}
\end{proposition}
\emph{Martin L\"{o}f -Solovay Chaitin random sequences obey the
Law of the Iterated Logarithm:}
\begin{equation}
    | \{ n \in \mathbb{N}_{+} \, : \,  \frac{\sum_{i=1}^{n} x_{i} }{\sqrt{2 n \log \log
    n}} \, \in  ( 1 - \epsilon , 1 + \epsilon ) \}| \; = \; \aleph_{0}  \; \; \forall \{ x_{n} \}_{n \in
    \mathbb{N}_{+}} \in  RANDOM_{MLSC} ( \{-1, 1
    \}^{\mathbb{N}_{+}}) ,  \forall \epsilon > 0
\end{equation}

From a foundational perspective one can think that the conceptual
corner stone of Probability Theory is the following:

\begin{proposition} \label{prop:algorithmic foundation of Probability Theory}
\end{proposition}
\emph{Algorithmic Foundation of Probability Theory:}

\begin{hypothesis}
\end{hypothesis}
\begin{center}
  $ \{ X_{n} \}_{n \in \mathbb{N}_{+} } $ independent identically
  distributed Bernoulli($\frac{1}{2}$) random variables on a
  binary alphabet A
\end{center}
\begin{center}
  $ x_{n} $ experimental value of $ X_{n} $
\end{center}

\begin{thesis}
\end{thesis}
\begin{equation}
    \{ x_{n} \}_{n \in \mathbb{N}_{+}} \in  RANDOM_{MLSC} ( A^{\mathbb{N}_{+}} )
\end{equation}

\begin{remark}
\end{remark}

Let us remark that the validity of the proposition \ref{prop:law
of the iterated logarithm} may be seen as a consequence of the
combination  of the Axiom of Reduction of Quantum Mechanics, i.e.
the axiom \ref{ax:Axiom of Reduction of Quantum Mechanics}, of the
proposition \ref{prop:algorithmic foundation of Probability
Theory} and of the proposition \ref{prop:Martin Lof-Solovay
Chaitin random sequences obey the Law of the Iterated Logarithm}.

\newpage
\section{Von Mises - Church randomness and its game theoretic meaning}

Given a binary alphabet $ A := \{ a_{1} , a_{2} \} $ and a partial
recursive map $ f : A^{+} \stackrel{\circ}{\mapsto} A $ let us
introduce the following:

\begin{definition}
\end{definition}
\emph{extraction map associated to f:}
\begin{equation}
    Extraction[f] : A^{\mathbb{N}_{+}} \mapsto A^{\mathbb{N}_{+}}
    \; : \;   Extraction[f] ( \bar{x} ) \; := \; \text{ ordered concatenation
    } (\{ x_{n} \, : \, f( x_{1}, \cdots , x_{n-1} ) = a_{2} \})
\end{equation}

We can then introduce the following \cite{von-Mises-81},
\cite{Van-Lambalgen-87a}, \cite{Van-Lambalgen-87b},
\cite{Van-Lambalgen-90}, \cite{Uspensky-Semenov-93},
\cite{Li-Vitanyi-97}:
\begin{definition} \label{def:Von Mises - Church random sequences}
\end{definition}
\emph{Von Mises - Church random sequences over A:}
\begin{multline}
    RANDOM_{VMC} ( A^{\mathbb{N}_{+}} ) \; := \; \{ \bar{x}
    \in A^{\mathbb{N}_{+}} \, : \,
     \lim_{n \rightarrow +
    \infty} \frac{N_{a}( \overrightarrow{Extraction[f] ( \bar{x} )}
(n))} {n} \; = \; \lim_{n \rightarrow +
    \infty} \frac{N_{a}( \vec{x} (n))}{n} \; = \; \frac{1}{2} \\
     \forall a \in A ,
    \forall \text{ partial
recursive map }  f : A^{+} \stackrel{\circ}{\mapsto} A \}
\end{multline}
where $ N_{a} ( \vec{x} ) $ is the number of times the letter a
occurs in the string $ \vec{x}$.

In the mentioned literature it is proved that:

\begin{proposition} \label{prop:Von Mises Church randomness is weaker than Martin Lof -
Solovay - Chaitin randomness}
\end{proposition}
\emph{Von Mises - Church randomness is weaker than Martin L\"{o}f
- Solovay - Chaitin randomness:}
\begin{equation}\
    RANDOM_{VMC} ( A^{\mathbb{N}_{+}} ) \; \supset \; RANDOM_{MLSC} ( A^{\mathbb{N}_{+}} )
\end{equation}

\begin{proposition} \label{prop:There exist Von Mises - Church random sequences violating
the Law of the Iterated Logarithm}
\end{proposition}
\emph{There exist Von Mises - Church random sequences violating
the Law of the Iterated Logarithm:}
\begin{equation}
 \exists \bar{x} \in     RANDOM_{VMC} ( \{ - 1 , + 1 \}^{\mathbb{N}_{+}} ) , \exists \epsilon \geq 0  \; :  \;  | \{ n \in \mathbb{N}_{+} \, : \,  \frac{\sum_{i=1}^{n} x_{i} }{\sqrt{2 n \log \log
    n}} \, \in \, ( 1 - \epsilon , 1 + \epsilon ) \}| \; < \; \aleph_{0}
\end{equation}

The notion of Von Mises - Church randomness has a natural game
theoretic meaning whose explanation requires the introduction of
some rudiments of Game Theory.

With this regard let us recall that \cite{Aumann-00},
\cite{Von-Neumann-Morgenstern-04}, \cite{ Nash-97a},
\cite{Nash-97b}, \cite{Nash-02}, \cite{Osborne-Rubinstein-94}:

\begin{definition} \label{def:strategic game}
\end{definition}
\emph{strategic game:}

a triple $ ( \mathcal{P} , \{ A_{i} \}_{i \in \mathcal{P}} ,  \{
\succeq_{i} \}_{i \in \mathcal{P}}   ) $ such that:
\begin{enumerate}
    \item $ \mathcal{P} $ is a finite set (the \emph{set of the
    players})
    \item  $ A_{i} $ is a non
    empty set (the \emph{set of the actions}
    available to the player i)
    \item $  \succeq_{i}  $ is a
    \emph{preference relation} on $ \times_{i \in \mathcal{P} }
    A_{i}$ (the \emph{preference relation} of the player i)
\end{enumerate}
where, given a set S:
\begin{definition}
\end{definition}
\emph{preference relation over S:}

a binary relation $ \succeq $ over S that is:
\begin{enumerate}
    \item complete, i.e.:
\begin{equation}
    x \succeq y \; \vee \;  y \succeq x \; \; \forall x , y \in S
\end{equation}
    \item reflexive, i.e.:
\begin{equation}
   x \succeq x \; \; \forall x \in S
\end{equation}
    \item transitive, i.e.:
\begin{equation}
   ( x \succeq y \, \wedge \, y \succeq z \; \Rightarrow \; x
    \succeq z ) \; \; \forall x , y , z \in S
\end{equation}
\end{enumerate}
($ \vee $ being the logical disjunction, i.e. the logical \emph{or
operator}).

\smallskip

Given a strategic game $ \mathcal{G} =    ( \mathcal{P} , \{ A_{i}
\}_{i \in \mathcal{P}} ,  \{ \succeq_{i} \}_{i \in \mathcal{P}} )
$, an element $  a = \{ a_{i} \}_{i \in \mathcal{P}} \in \times_{i
\in \mathcal{P} } A_{i} $ is called a \emph{profile of actions} of
$ \mathcal{G} $.

The abstract mathematical notion of a preference relation may be
often described by a more concrete utility function, according to
the following:
\begin{definition} \label{def:utilility functions of a strategic games}
\end{definition}
\emph{utility functions for $\mathcal{G} $:}

a set of maps $ \{ u_{i} \}_{i \in \mathcal{P}} $ such that:
\begin{equation}
    u_{i} :   \times_{i \in \mathcal{P} } A_{i} \mapsto
    \mathbb{R} \; \; \forall i \in \mathcal{P}
\end{equation}
\begin{equation}
    u_{i} (a) \geq u_{i} (b) \; \Leftrightarrow \; a
    \succeq_{i} b \; \; \forall a , b \in \times_{i \in \mathcal{P} } A_{i}
\end{equation}
Clearly to assign \emph{utility functions} of a \emph{strategic
game} is equivalent to assign its \emph{preference relations}.

Let us then introduce the following fundamental:
\begin{definition}
\end{definition}
\emph{Nash equilibrium of $\mathcal{G} $:}

a \emph{profile of actions} $ a^{\star} = \{ a_{i}^{\star} \}_{i
\in \mathcal{P}} \in \times_{i \in \mathcal{P} } A_{i} $ such
that:
\begin{equation}
    a^{\star}   \succeq_{i} \{ a_{j}^{\star} , a_{i} \}_{j \in \mathcal{P} -
   \{ i \} } \; \; \forall a_{i} \in A_{i}  ,
    \forall i \in \mathcal{P}
\end{equation}

\smallskip

We will denote by $ \mathcal{E}_{Nash}( \mathcal{G} ) $ the set of
the Nash equilibriums of the strategic game $  \mathcal{G} $.

\smallskip

Given a strategic game with 2 players $ \mathcal{G} := ( \{ Alice
, Bob \} , \{ A_{Alice} , A_{Bob} \} , \{ \succeq_{Alice} ,
\succeq_{Bob} \} ) $:
\begin{definition}
\end{definition}
\emph{$ \mathcal{G}$ is strictly competitive:}
\begin{equation}
 x \succeq_{Alice} y \; \Leftrightarrow \; y \succeq_{Bob} x \; \;
 \forall x,y \in A_{Alice} \times A_{Bob}
\end{equation}

Given a strictly competitive game  $ \mathcal{G} = ( \{ Alice ,
Bob \} \, , \, \{ A_{Alice} , A_{Bob} \} \, , \,  \{
\succeq_{Alice} , \succeq_{Bob} \} ) $ let $ \{ u_{Alice} ,
u_{Bob} \} $ be \emph{utiliity functions} of $ \mathcal{G} $.

Without loss of generality we may assume that:
\begin{equation}
    u_{Bob} ( x , y ) \; := - \; u_{Alice} ( x,y) \; \; \forall x \in
    A_{Alice} ,  \forall y \in A_{Bob}
\end{equation}

Given an action $ x^{\star} \in A_{Alice} $:
\begin{definition}
\end{definition}
\emph{$ x^{\star} $ is a maxminimizer for Alice:}
\begin{equation}
    \min_{y \in A_{Bob}} u_{Alice} (  x^{\star} , y) \; \geq \; \min_{y \in A_{Bob}} u_{Alice} (  x ,
    y) \; \; \forall x \in A_{Alice}
\end{equation}
Similarly, given an action $ y^{\star} \in A_{Bob} $:
\begin{definition}
\end{definition}
\emph{$ y^{\star} $ is a maxminimizer for Bob:}
\begin{equation}
    \min_{x \in A_{Alice}} u_{Bob} (  x , y^{\star}) \; \geq \; \min_{x \in A_{Alice}} u_{Bob} (  x ,
    y) \; \; \forall y \in A_{Bob}
\end{equation}

The following fundamental theorem shows how the Von Neumann and
Morgenstern's theory of strictly competitive stategic games is a
particular case of Nash's theory:

\begin{proposition} \label{prop:About Nash equilibria of a strictly competitive strategic
game}
\end{proposition}
\emph{About Nash equilibriums of a strictly competitive strategic
game:}
\begin{enumerate}
    \item if $ ( x^{\star} , y^{\star} ) $ is a Nash equilibrium
    of $ \mathcal{G} $ then $  x^{\star} $ is a maximinimer for
    Alice and $  y^{\star} $ is a maxminimizer for Bob.
    \item if $ ( x^{\star} , y^{\star} ) $ is a Nash equilibrium
    of $ \mathcal{G} $ then $ \max_{x \in A_{Alice}} \min_{y \in
    A_{Bob}} u_{Alice} ( x ,y ) = \min_{y \in
    A_{Bob}} \max_{x \in A_{Alice}} u_{Alice} ( x ,y ) =  u_{Alice} ( x^{\star} ,y^{\star}
    ) $.
    \item if $ \max_{x \in A_{Alice}} \min_{y \in
    A_{Bob}} u_{Alice} ( x ,y ) = \min_{y \in
    A_{Bob}} \max_{x \in A_{Alice}} u_{Alice} ( x ,y ) $, $
    x^{\star}$ is a maximinimizer for Alice and $
    y^{\star}$  is a maximinimizer for Bob, then $ (  x^{\star} ,
    y^{\star})$ is a Nash equilibrium of $ \mathcal{G} $.
\end{enumerate}

\bigskip

Let us now consider the following game between Alice and Bob:

at the beginning of the game Alice chooses a sequence $ \{ x_{n}
\}_{n \in \mathbb{N}_{+}} \in \{ -1 , + 1 \}^{\mathbb{N}_{+}} $
while Bob chooses a partial recursive function $ f : \{ -1 , 1
\}^{+} \stackrel{\circ}{\mapsto} \{ -1 , 1 \} $ encoding his
strategy:

Then, at the $ n^{th} $ turn:
\begin{enumerate}
    \item Alice communicates to Bob the bit $x_{n-1} $
    \item Bob bets on the value of the bit $ x_{n} $ according to
    the following algorithm:
\begin{equation}
    betting(n) \; := \; \left\{%
\begin{array}{ll}
    f( x_{1}, \cdots , x_{n-1} ) & \hbox{if $ ( x_{1} , \cdots , x_{n-1} ) \in HALT_{f}$} \\
    0, & \hbox{ otherwise.} \\
\end{array}%
\right.
\end{equation}
using the information available to him, namely the value of the
bits $ x_{1} , \cdots , x_{n-1} $, and his strategy f.
\end{enumerate}

The payoffs gained by Alice and Bob at the $ n^{th} $ turn are
given by the following rules:
\begin{equation}
    payoff_{Alice} (n) \; := \left\{%
\begin{array}{ll}
    0, & \hbox{if $betting(n) = 0 $} \\
    +1, & \hbox{if $ betting(n) \neq 0 $ and $ x_{n} \neq  betting(n)$} \\
    -1, & \hbox{if $ betting(n) \neq 0 $ and $ x_{n} =  betting(n)$} \\
\end{array}%
\right.
\end{equation}
\begin{equation}
    payoff_{Bob} (n) \; := \left\{%
\begin{array}{ll}
    0, & \hbox{if $betting(n) = 0 $;} \\
    +1, & \hbox{if $ betting(n) \neq 0 $ and $ x_{n} =  betting(n)$} \\
    -1, & \hbox{if $ betting(n) \neq 0 $ and $ x_{n} \neq  betting(n)$} \\
\end{array}%
\right.
\end{equation}

The corner-stone of the Von Mises - Church definition of
randomness  is the following:
\begin{proposition} \label{prop:Fundamental theorem about Von Mises - Church randomness}
\end{proposition}
\emph{Fundamental theorem about Von Mises - Church randomness:}
\begin{multline}
    \sum_{n=1}^{\infty} payoff_{Alice} (n) \; = \; \sum_{n=1}^{\infty} payoff_{Bob}
    (n) \, = \, 0 \; \; \forall  f : \{ -1 , 1 \}^{+} \stackrel{\circ}{\mapsto} \{ -1 , 1
\} \, \text{ partial recursive } \; \\
\Updownarrow \\
\{ x_{n} \}_{n \in \mathbb{N}_{+}} \in RANDOM_{VMC} ( \{ -1 , +1
\}^{\mathbb{N}_{+}})
\end{multline}
\begin{proof}
Introduced:
\begin{equation}
   bias(\bar{x} , f )  \; : = \;  \lim_{n \rightarrow +
    \infty} \frac{N_{a_{2}}( \overrightarrow{Extraction[f] ( \bar{x} )}
(n))} {n}  - \frac{1}{2}
\end{equation}
the condition:
\begin{equation}
    bias(\bar{x} , f )  \; = \; 0  \; \; \forall  f : \{ -1 , 1 \}^{+} \stackrel{\circ}{\mapsto} \{ -1 , 1
\} \, \text{ partial recursive }
\end{equation}
is equivalent to the condition that there is no regularity in $
\bar{x} $ of which Bob can take advantage of in  choosing his
strategy f so that $ \sum_{n=1}^{\infty} payoff_{Bob} (n) > 0 $.

Taking also into account that obviously:
\begin{equation}
N_{a_{2}}( \overrightarrow{Extraction[f] ( \bar{x} )}(n)) \; = \;
n - N_{a_{1}}( \overrightarrow{Extraction[f] ( \bar{x} )} (n)) \;
\; \forall n \in \mathbb{N}_{+}
\end{equation}
and:
\begin{equation}
    payoff_{Alice} (n) \; = \; -  \, payoff_{Bob} (n)  \; \; \forall
n \in \mathbb{N}_{+}
\end{equation}
the thesis follows.
\end{proof}

\bigskip

The proposition \ref{prop:Fundamental theorem about Von Mises -
Church randomness} states that in the involved game there are no
winners, whichever is the strategy of Bob, if and only if the
sequence initially chosen by Alice is Von Mises - Church random.

It appears then natural to restate it mathematically in
game-theoretic words in terms of the Nash equilibriums of a
strategic game (according to the definition \ref{def:strategic
game}) formalizing the stuff that I have described in informal
words.

Let us, with this regard, introduce the following:
\begin{definition}
\end{definition}
\emph{Von Mises - Church game:}

the strategic game  $ \mathcal{G}_{VMC} :=  ( \mathcal{P} , \{
A_{i} \}_{i \in \mathcal{P}} , \{ \succeq_{i} \}_{i \in
\mathcal{P}}   ) $ such that:
\begin{equation}
    \mathcal{P} \; := \{ Alice , Bob \}
\end{equation}
\begin{equation}
    A_{Alice} \; := \; \{ -1 , + 1 \}^{\mathbb{N}_{+}}
\end{equation}
\begin{equation}\
    A_{Bob} \; := \;  \{ f : \{ -1 , 1 \}^{+} \stackrel{\circ}{\mapsto} \{ -1 , 1
\} \, \text{ partial recursive } \}
\end{equation}
\begin{multline}
  \{  u_{Alice}( a_{Alice} , a_{Bob} )  :=  \sum_{n=1}^{\infty}
   payoff_{Alice}(n) \: , \: u_{Bob}( a_{Alice} , a_{Bob} ) \; :=  \sum_{n=1}^{\infty}
   payoff_{Bob}(n) ( a_{Alice} , a_{Bob} ) \} \\
   \text{ are utility functions of $ \mathcal{G}_{VMC}$}
\end{multline}

Then:
\begin{proposition} \label{prop:Von Mises Church randomness versus Nash equilibrium}
\end{proposition}
\emph{Von Mises - Church randomness versus Nash equilibrium:}
\begin{equation}
    \mathcal{E}_{Nash} (  \mathcal{G}_{VMC} ) \; \subseteq \; RANDOM_{VMC} ( \{ -1 , +1
\}^{\mathbb{N}_{+}}) \times A_{Bob}
\end{equation}
\begin{proof}
Given $ a_{Alice}^{\star} \in RANDOM_{VMC} ( \{ -1 , +1
\}^{\mathbb{N}_{+}}) $ the proposition \ref{prop:Fundamental
theorem about Von Mises - Church randomness} implies that:
\begin{equation}
    \forall a_{Alice} \in ( \{ -1 , +1
\}^{\mathbb{N}_{+}} - RANDOM_{VMC} ( \{ -1 , +1
\}^{\mathbb{N}_{+}})) \; \exists a_{Bob}^{\star} \in A_{bob} \; :
\\
\; (  a_{Alice}^{\star} , a_{Bob}^{\star} ) \succeq_{Alice} (
a_{Alice} ,  a_{Bob}^{\star} )
\end{equation}
Since clearly the Von Mises - Church game $ \mathcal{G}_{VMC} $
 is strictly competitive, the thesis follows applying the proposition
\ref{prop:About Nash equilibria of a strictly competitive
strategic game} to $ \mathcal{G}_{VMC} $.

\end{proof}

\smallskip

\begin{remark}
\end{remark}
Let us remark that the definition \ref{def:Von Mises - Church
random sequences} is different (though  equivalent) to the
traditional one in that the admissible place selection rules have
been allowed to be \emph{partial} maps.

I have chosen such a formulation since it emphasizes how, under
the assumption of the \emph{Church-Turing Thesis}
\cite{Davis-94a}, \cite{Davis-94b}, \cite{Davis-04}, the set of
Bob's actions in the \emph{Von Mises - Church game} is exactly the
set of available strategies that Bob can \emph{effectively}
pursue.

Such a rule of Computability Theory in delimiting the strategies
that can be effectively performed generates a bridge between Game
Theory and Theoretical Computer Science whose conceptual depth
begins at last to be appreciated (see for instance
\cite{Tardos-Vazirani-07}, \cite{Papadimitriou-07}).

\newpage
\section{Experimental Iterated Logarithm randomness test  not passed by the quantum coin} \label{sec:Experimental violation of the Iterated Logarithm
randomness test}

The first step in the strategic planning of an experiment aiming
at performing the Iterated Logarithm randomness test on a
statistically significative sample of independent tosses of a
quantum coin has been, since at the moment I can only to
count on my own private resources, the purchase of the quantum
random number generator \emph{Quantis} realized by the spin-off
company of the University of Geneva \emph{id Quantique}
\cite{idQuantique-08} (based on the simplest beam splitter, i.e. a
semi-trasparent mirror \cite{Leonhardt-97}; for the credentials of
\emph{Quantis} see
\cite{Jennewein-Achleitner-Weichs-Weinfurter-Zeilinger-00},
\cite{randomnumbers-07} and \cite{Calude-05}).

The results contained on this paper are, consequentially, based on
the assumption that \emph{Quantis} is a fair quantum coin.

I have then used \emph{Quantis} to generate $10^2$ binary strings,
each of length $ 10^4 $, that are reported in the appendix
\ref{sec:Experimental data}.

Let us now analyze these data.

Introduced the quantity:
\begin{equation}
    s_{n} \; := \; \sum_{i=1}^{n} x_{i}
\end{equation}
its value is clearly given by the sum of the tosses giving as
result "+ 1" minus the number of the tosses giving as result "- 1
":
\begin{equation}
    s_{n} \; = N_{1}[\vec{x}(n)] \, - \, ( n - N_{1}[\vec{x}(n)])
\end{equation}
and hence:
\begin{equation}
    s_{n} \; = \; 2  N_{1}[\vec{x}(n)]  - n
\end{equation}
Let us now observe that:
\begin{equation}
    s_{n} \, = \, k \; \Leftrightarrow \; 2  N_{1}[\vec{x}(n)]  -
    n \, = \, k \; \Leftrightarrow \;  N_{1}[\vec{x}(n)] \, = \,
    \frac{n+k}{2}  \; \; \forall k \in \{ - n , \cdots , n \} : (
    -1)^{n+k} = 1
\end{equation}
Hence:
\begin{equation} \label{eq:probability distribution of the random walk over the integers}
    P ( s_{n} = k) \; = \; \delta_{ (
    -1)^{n+k},1} \, P (  N_{1}[\vec{x}(n)] \, = \,
    \frac{n+k}{2} ) \; = \;  \delta_{ (
    -1)^{n+k},1}  \binom{n}{\frac{n+k}{2}} \frac{1}{2^{n}}
\end{equation}
where $ \delta_{x,y} \; := \; \left\{%
\begin{array}{ll}
    1, & \hbox{if $ x = y $} \\
    0, & \hbox{otherwise} \\
\end{array}%
\right. $ is the usual Kronecker delta.

Let us now introduce the events:
\begin{equation}
    E_{n,\epsilon} \; := \; \text{ the sentence  }
    \frac{s_{n}}{\phi(n)} \in [ 1 - \epsilon , 1 +
    \epsilon ] \text{ is true }
\end{equation}
where:
\begin{equation}
    \phi (n) \; := \; \sqrt{ 2 n \log \log n }
\end{equation}

 Let us remark that:
\begin{equation} \label{eq:cases one and two}
    P(E_{1,\epsilon} ) \; = \; P(E_{2,\epsilon} ) \; = \; 0
\end{equation}
since in these case the sentence $ \frac{s_{n}}{\phi(n)} \in [ 1 -
\epsilon , 1 + \epsilon ] $ is meaningless and hence, in
particular, it is not true.

Clearly:

\begin{multline} \label{eq:elementary probability}
    P(  E_{n , \epsilon} ) \; = \; \sum_{k=- n}^{+ n}  \theta [ k
    -  \phi (n)  ( 1 -
    \epsilon ) ] \theta [  \phi (n)  ( 1 +
    \epsilon ) - k ] P ( s_{n} = k ) \; = \\
     \sum_{k=- n}^{+ n} \theta [
    k  -  \phi (n) ( 1 -
    \epsilon ) ]   \theta [  \phi (n)  ( 1 +
    \epsilon) - k  ] \delta_{ (
    -1)^{n+k},1}  \binom{n}{\frac{n+k}{2}} \frac{1}{2^{n}}
\end{multline}
where  $ \theta(x) \; := \; \left\{%
\begin{array}{ll}
    1, & \hbox{if $ x \geq 0$} \\
    0, & \hbox{otherwise} \\
\end{array}%
\right.         $ is the Heaviside function \footnote{Let us
remark that I have defined the Heaviside function in a way such
that:
\begin{equation}
    \theta ( x  + \sqrt{ 2 \log \log 1  }) \; = \; \theta ( x + \sqrt{ 4 \log \log 2  } ) \; = \; 0 \; \; \forall x \in \mathbb{R}
\end{equation}
}.

Let us now introduce the following:
\begin{definition}
\end{definition}
\emph{counter random variable:}
\begin{equation}\label{eq:counter random variable}
    c_{N, \epsilon} \; := \; | \{ n \in \{ 1 , \cdots , N \} \, : \,
    E_{n,\epsilon} \} |
\end{equation}

Clearly $ c_{N, \epsilon} $ takes values in the set $ \{ 0 ,
\cdots , N \} $.

\begin{enumerate}

 \item Let us consider first of all the case $ m \neq 0 $. Then:

\begin{equation}
    P(c_{N, \epsilon} = m ) \; = \; \sum_{\pi \subset \{ 3 , \cdots , N \} : |\pi | = m
    } P ( \wedge_{i=1}^{m} E_{\pi(i), \epsilon} ) \; \; \forall m
    \in \{ 1 , \cdots , N \}
\end{equation}
where $ \wedge $ is the logical conjunction and where $ \pi = \{
\pi (1) , \cdots, \pi (m) \} $ is such that:
\begin{equation}
    \pi(i) \; < \;   \pi(j) \; \; \forall i < j
\end{equation}

Let us now observe that:
\begin{equation}
 P ( \wedge_{i=1}^{m}  E_{ \pi(i) , \epsilon} ) \; = \; P(  E_{\pi(1),\epsilon} )
 \prod_{i=2}^{m} P( E_{ \pi(i) , \epsilon} | \wedge_{j=1}^{i-1}  E_{\pi(j) , \epsilon} )
\end{equation}

Let us then remark that the markovianity of the underlying random
walk over the integers implies that:
\begin{equation}
 P ( \wedge_{i=1}^{m}  E_{ \pi(i) , \epsilon} ) \; = \; P(  E_{\pi(1),\epsilon} )
 \prod_{i=2}^{m} P( E_{ \pi(i) , \epsilon} |   E_{\pi(i-1) , \epsilon} )
\end{equation}
where:
\begin{equation} \label{eq:two cases}
  P( E_{ \pi(i) , \epsilon} |   E_{\pi(i-1) , \epsilon} ) \; = \; \left\{%
\begin{array}{ll}
    P( E_{ \pi(i) , \epsilon} |   E_{\pi(i)-1 , \epsilon} ) & \hbox{if $ \pi(i-1) = \pi(i) -1 $} \\
    P( E_{ \pi(i) , \epsilon} )  & \hbox{if $  \pi(i-1) < \pi(i) -1 $} \\
\end{array}%
\right.
\end{equation}

Hence:
\begin{multline}
    P(c_{N, \epsilon} = m ) \; = \; \sum_{\pi \subset \{ 3 , \cdots , N \} : |\pi | = m
    } P(  E_{\pi(1),\epsilon} )  \prod_{i=2}^{m} \{ \delta_{\pi(i-1), \pi(i) -1
    }  P( E_{ \pi(i) , \epsilon} |   E_{\pi(i)-1 , \epsilon} )  +
    \\
    \theta [ -1 - \pi(i-1) + \pi(i) ] ( 1 - \delta_{-1 - \pi(i-1)
    +  \pi(i),0} ) P( E_{ \pi(i) , \epsilon} ) \} \; \; \forall m \in
    \{ 1 , \cdots , N \}
\end{multline}

Introduced the maps:
\begin{equation}
    f( n , \epsilon ) \; := \; [  \phi (n) -  \phi (n-1) ] ( 1 -
    \epsilon )
\end{equation}
\begin{equation}
    g( n, \epsilon ) \; := \;  [  \phi (n) -  \phi (n-1) ] ( 1 +
    \epsilon )
\end{equation}
let us remark that $ f( n, \epsilon) $ and $ g( n, \epsilon) $
take real values if and only the integer variable $ n \geq 4  $
and that:
\begin{equation}
 0 < f ( n , \epsilon ) < g ( n , \epsilon ) < 1 \; \; \forall n
 \geq 4 , \forall \epsilon \leq 0.1
\end{equation}

Let us furthermore introduce the  following random variables:
\begin{equation}
   \delta_{L,n,\epsilon} \; := \; s_{n} -  \phi (n) ( 1 - \epsilon)
\end{equation}
\begin{equation}
   \delta_{R,n,\epsilon} \; := \;\phi (n) ( 1 + \epsilon ) - s_{n}
\end{equation}

Considering the inequalities obtained assuming the conditional
hypothesis $ E_{\pi(i)-1 , \epsilon}  $ and taking into account
the possible outcomes of  $  x_{\pi(i)} $ it follows that:
\begin{enumerate}
    \item if $ x_{\pi(i)} = - 1 $ then:
\begin{equation}
 [ \delta_{L, \pi(i)-1 ,\epsilon} \geq 0 \, \Rightarrow \, \delta_{L, \pi(i)
 ,\epsilon} \geq 0 ] \; \Leftrightarrow \; \delta_{L, \pi(i)-1
 ,\epsilon}  - f[ \pi(i) , \epsilon ] -1 \geq 0
\end{equation}
\begin{equation}
   \delta_{R, \pi(i)-1 ,\epsilon} \geq 0 \; \Rightarrow \; \delta_{R, \pi(i)
 ,\epsilon} \geq 0
\end{equation}
  \item if $ x_{\pi(i)} = + 1 $ then:
\begin{equation}
   \delta_{L, \pi(i)-1 ,\epsilon} \geq 0 \; \Rightarrow \; \delta_{L, \pi(i)
 ,\epsilon} \geq 0
\end{equation}
\begin{equation}
 [ \delta_{R, \pi(i)-1 ,\epsilon} \geq 0 \, \Rightarrow \, \delta_{R, \pi(i)
 ,\epsilon} \geq 0 ] \; \Leftrightarrow \; \delta_{R, \pi(i)-1
 ,\epsilon}  + g[ \pi(i) , \epsilon ] -1 \geq 0
\end{equation}
\end{enumerate}

Therefore:
\begin{multline}
    P( E_{ \pi(i) , \epsilon} |   E_{\pi(i)-1 , \epsilon} ) \; =
    \; \frac{1}{2} P ( \delta_{L, \pi(i)-1
 ,\epsilon}  - f[ \pi(i) , \epsilon ] - 1 \geq 0 ) + \frac{1}{2} P
 ( \delta_{R, \pi(i)-1
 ,\epsilon}  + g[ \pi(i) , \epsilon ] -1 \geq 0 ) \; = \\
 \frac{1}{2} \sum_{k= - ( \pi(i) -1 ) }^{\pi(i) -1} \theta \{ k -
 \phi[ \pi(i) -1 ] ( 1 - \epsilon ) - f [ \pi(i) , \epsilon ] - 1
 \} P ( s_{\pi(i) -1 } = k ) +  \\
 \frac{1}{2} \sum_{k= - ( \pi(i) -1 ) }^{\pi(i) -1} \theta \{
 \phi[ \pi(i) -1 ] ( 1 + \epsilon ) - k + g [ \pi(i) , \epsilon ] - 1
 \} P ( s_{\pi(i) -1 } = k )
\end{multline}
that using the equation \ref{eq:probability distribution of the
random walk over the integers} gives:
\begin{multline} \label{eq:conditional probability}
    P( E_{ \pi(i) , \epsilon} |   E_{\pi(i)-1 , \epsilon} ) \; =
 \frac{1}{2} \sum_{k= - ( \pi(i) -1 ) }^{\pi(i) -1} \theta \{ k -
 \phi[ \pi(i) -1 ] ( 1 - \epsilon ) - f [ \pi(i) , \epsilon ] - 1
 \} \delta_{ (
    -1)^{\pi(i)-1+k},1}  \binom{\pi(i)-1}{\frac{\pi(i)-1+k}{2}} \frac{1}{2^{\pi(i)-1}}  + \\
  \frac{1}{2} \sum_{k= - ( \pi(i) -1 ) }^{\pi(i) -1} \theta \{
 \phi[ \pi(i) -1 ] ( 1 + \epsilon ) - k + g [ \pi(i) , \epsilon ] - 1
 \} \delta_{ (
    -1)^{\pi(i)-1+k},1}  \binom{\pi(i)-1}{\frac{\pi(i)-1+k}{2}} \frac{1}{2^{\pi(i)-1}}
\end{multline}

Hence:
\begin{multline} \label{eq:counter distribution for nonzero m}
  P(c_{N, \epsilon} = m ) \; = \; \sum_{\pi \subset \{ 3 , \cdots , N \} : |\pi | = m
    }  \sum_{k=-  \pi(1)  }^{\pi(1) } \theta [
    k  -  \phi (\pi(1) ) ( 1 -
    \epsilon ) ]   \theta [  \phi (\pi(1) )  ( 1 +
    \epsilon) - k  ] \cdot
    \delta_{ (-1)^{\pi(1) +k},1}  \binom{n}{\frac{\pi(1) +k}{2}}
\frac{1}{2^{\pi(1) }} \cdot
     \\ \prod_{i=2}^{m} \{ \delta_{\pi(i-1), \pi(i) -1
    } [  \frac{1}{2} \sum_{k= - ( \pi(i) -1 ) }^{\pi(i) -1} \theta \{ k -
 \phi[ \pi(i) -1 ] ( 1 - \epsilon ) - f [ \pi(i) , \epsilon ] - 1
 \} \delta_{ (
    -1)^{\pi(i)-1+k},1}  \binom{\pi(i)-1}{\frac{\pi(i)-1+k}{2}} \frac{1}{2^{\pi(i)-1}}  + \\
  \frac{1}{2} \sum_{k= - ( \pi(i) -1 ) }^{\pi(i) -1} \theta \{
 \phi[ \pi(i) -1 ] ( 1 + \epsilon ) - k + g [ \pi(i) , \epsilon ] - 1
 \} \delta_{ (
    -1)^{\pi(i)-1+k},1}  \binom{\pi(i)-1}{\frac{\pi(i)-1+k}{2}} \frac{1}{2^{\pi(i)-1}} ] +
    \\
    \theta [ -1 - \pi(i-1) + \pi(i) ] ( 1 - \delta_{-1 - \pi(i-1)
    +
    \pi(i),0} ) \sum_{k=-  \pi(i)  }^{\pi(i) } \theta [
    k  -  \phi (\pi(i) ) ( 1 -
    \epsilon ) ]   \theta [  \phi (\pi(i) )  ( 1 +
    \epsilon) - k  ]
    \delta_{ (-1)^{\pi(i) +k},1}  \binom{n}{\frac{\pi(i) +k}{2}}
\frac{1}{2^{\pi(i) }} \} \\  \forall m \in \{ 1 , \cdots N \}
\end{multline}

\item Let us now consider the case m=0:
\begin{equation}
    P(c_{N, \epsilon} = 0 ) \; = \;  P ( \wedge_{n=1}^{N}  \neg E_{n,\epsilon} ) \; = \; P( \neg E_{1,\epsilon} )
 \prod_{n=2}^{N} P(  \neg E_{n,\epsilon} | \wedge_{k=1}^{n-1} \neg E_{k,\epsilon} )
\end{equation}
(where $ \neg $ is the \emph{logical negation}, i.e. the logical
\emph{not operator}) that, by the equation \ref{eq:cases one and
two}, reduces to:
\begin{equation}
 P ( \wedge_{n=1}^{N}  \neg E_{n,\epsilon} ) \; = \;
 \prod_{n=3}^{N} P(  \neg E_{n,\epsilon} | \wedge_{k=3}^{n-1} \neg E_{k,\epsilon} )
\end{equation}

Let us then remark that the markovianity of the underlying random
walk over the integers implies that:
\begin{equation}
     P(  \neg E_{n,\epsilon} | \wedge_{k=3}^{n-1} \neg E_{k,\epsilon} ) \; = \;  P(  \neg E_{n,\epsilon} | \neg E_{n-1,\epsilon}
     ) \; \; \forall n \in \mathbb{N} : n \geq 3
\end{equation}
Let us observe that:
\begin{enumerate}
     \item if $ n = 3 $ then:
\begin{equation} \label{eq:case three}
    P(  \neg E_{3,\epsilon} | \wedge_{k=3}^{3} \neg E_{k,\epsilon}  ) \; = \; P ( \neg E_{3,\epsilon} )
    \; = 1 - P(E_{3,\epsilon}) \; = \; 1 -  \sum_{k=- 3}^{+ 3} \theta [
    k  -  \phi (3) ( 1 -
    \epsilon ) ]   \theta [  \phi (3)  ( 1 +
    \epsilon) - k  ] \delta_{ (
    -1)^{3+k},1}  \binom{3}{\frac{3+k}{2}} \frac{1}{2^{3}}
\end{equation}
where I have used the equation \ref{eq:cases one and two} and the
equation \ref{eq:elementary probability}
    \item if $ n > 3 $ then considering the inequalities obtained assuming the conditional
hypothesis $ \neg E_{n-1 , \epsilon}  $ and taking into accounts
the possible outcomes of  $  x_{n} $ it follows that:
    \begin{enumerate}
        \item if $ x_{n}= - 1 $ then:
\begin{equation}
    \delta_{L,n-1,\epsilon} < 0 \; \Rightarrow \;
    \delta_{L,n,\epsilon} < 0
\end{equation}
\begin{equation}
   [ \delta_{R,n-1,\epsilon} < 0 \, \Rightarrow \,
   \delta_{R,n,\epsilon} < 0 ] \; \Leftrightarrow \; \delta_{R,n-1,\epsilon}
   + g(n,\epsilon) + 1 < 0
\end{equation}
        \item if $ x_{n}= + 1 $ then:
\begin{equation}
   [ \delta_{L,n-1,\epsilon} < 0 \, \Rightarrow \,
   \delta_{L,n,\epsilon} < 0 ] \; \Leftrightarrow \; \delta_{L,n-1,\epsilon}
   - f(n,\epsilon) + 1 < 0
\end{equation}
\begin{equation}
    \delta_{R,n-1,\epsilon} < 0 \; \Rightarrow \;
    \delta_{R,n,\epsilon} < 0
\end{equation}
Therefore:
\begin{multline} \label{eq:conditional probability in the zero case}
    P(  \neg E_{n,\epsilon} | \neg E_{n-1,\epsilon}) \; = \;
    \frac{1}{2} P [ \delta_{R,n-1,\epsilon}
   + g(n,\epsilon) + 1 < 0 ] +  \frac{1}{2}  P [  \delta_{L,n-1,\epsilon}
   - f(n,\epsilon) + 1 < 0 ] \; = \\
   \frac{1}{2} \sum_{k=-(n-1)}^{n-1} \theta[ k - \phi(n-1) ( 1 +
   \epsilon) - g(n , \epsilon ) - 1 ] ( 1 - \delta_{k - \phi(n-1) ( 1 +
   \epsilon) - g(n , \epsilon ) - 1,0}) \delta_{ (
    -1)^{n-1+k},1}  \binom{n-1}{\frac{n-1+k}{2}} \frac{1}{2^{n-1}} + \\
   \frac{1}{2} \sum_{k=-(n-1)}^{n-1} \theta[ \phi(n-1) ( 1 -
   \epsilon) - k + f(n , \epsilon ) - 1 ] ( 1 - \delta_{ \phi(n-1) ( 1 -
   \epsilon) - k + f(n , \epsilon ) - 1 ,0}) \delta_{ (
    -1)^{n-1+k},1}  \binom{n-1}{\frac{n-1+k}{2}} \frac{1}{2^{n-1}}
\end{multline}
\end{enumerate}
\end{enumerate}

Hence:
\begin{multline}  \label{eq:counter distribution for zero m}
    P ( c_{N,\epsilon} = 0 ) \; = \; \{ 1 -  \sum_{k=- 3}^{+ 3} \theta [
    k  -  \phi (3) ( 1 -
    \epsilon ) ]   \theta [  \phi (3)  ( 1 +
    \epsilon) - k  ] \delta_{ (
    -1)^{3+k},1}  \binom{3}{\frac{3+k}{2}} \frac{1}{2^{3}} \}
    \cdot \\
    \prod_{n=4}^{N} \{ \frac{1}{2} \sum_{k=-(n-1)}^{n-1} \theta[ k - \phi(n-1) ( 1 +
   \epsilon) - g(n , \epsilon ) - 1 ] ( 1 - \delta_{k - \phi(n-1) ( 1 +
   \epsilon) - g(n , \epsilon ) - 1,0}) \delta_{ (
    -1)^{n-1+k},1}  \binom{n-1}{\frac{n-1+k}{2}} \frac{1}{2^{n-1}} + \\
   \frac{1}{2} \sum_{k=-(n-1)}^{n-1} \theta[ \phi(n-1) ( 1 -
   \epsilon) - k + f(n , \epsilon ) - 1 ] ( 1 - \delta_{ \phi(n-1) ( 1 -
   \epsilon) - k + f(n , \epsilon ) - 1 ,0}) \delta_{ (
    -1)^{n-1+k},1}  \binom{n-1}{\frac{n-1+k}{2}} \frac{1}{2^{n-1}} \}
\end{multline}
\end{enumerate}

Substituting the equation \ref{eq:counter distribution for zero m}
and the equation \ref{eq:counter distribution for nonzero m} into
the equation:
\begin{equation}
    P ( c_{N,\epsilon} = m ) \; = \; \delta_{m,0} P ( c_{N,\epsilon} = 0
    )  \, + \, \theta(m) ( 1 - \delta_{m,0}) P ( c_{N,\epsilon} = m > 0
    ) \; \; \forall m \in \{ 0 , \cdots , N \}
\end{equation}

one obtains the general formula for the probability distribution
of the counter variable:
\begin{multline} \label{eq:counter distribution}
  P ( c_{N,\epsilon} = m ) \; = \; \delta_{m,0} \{ 1 -  \sum_{k=- 3}^{+ 3} \theta [
    k  -  \phi (3) ( 1 -
    \epsilon ) ]   \theta [  \phi (3)  ( 1 +
    \epsilon) - k  ] \delta_{ (
    -1)^{3+k},1}  \binom{3}{\frac{3+k}{2}} \frac{1}{2^{3}} \}
    \cdot \\
    \prod_{n=4}^{N} \{ \frac{1}{2} \sum_{k=-(n-1)}^{n-1} \theta[ k - \phi(n-1) ( 1 +
   \epsilon) - g(n , \epsilon ) - 1 ] ( 1 - \delta_{k - \phi(n-1) ( 1 +
   \epsilon) - g(n , \epsilon ) - 1,0}) \delta_{ (
    -1)^{n-1+k},1}  \binom{n-1}{\frac{n-1+k}{2}} \frac{1}{2^{n-1}} + \\
   \frac{1}{2} \sum_{k=-(n-1)}^{n-1} \theta[ \phi(n-1) ( 1 -
   \epsilon) - k + f(n , \epsilon ) - 1 ] ( 1 - \delta_{ \phi(n-1) ( 1 -
   \epsilon) - k + f(n , \epsilon ) - 1 ,0}) \delta_{ (
    -1)^{n-1+k},1}  \binom{n-1}{\frac{n-1+k}{2}} \frac{1}{2^{n-1}} \}  \, + \\
     \theta(m) ( 1 - \delta_{m,0})  \sum_{\pi \subset \{ 3 , \cdots , N \} : |\pi | = m
    }  \sum_{k=-  \pi(1)  }^{\pi(1) } \theta [
    k  -  \phi (\pi(1) ) ( 1 -
    \epsilon ) ]   \theta [  \phi (\pi(1) )  ( 1 +
    \epsilon) - k  ] \cdot
    \delta_{ (-1)^{\pi(1) +k},1}  \binom{n}{\frac{\pi(1) +k}{2}}
\frac{1}{2^{\pi(1) }} \cdot
     \\ \prod_{i=2}^{m} \{ \delta_{\pi(i-1), \pi(i) -1
    } [  \frac{1}{2} \sum_{k= - ( \pi(i) -1 ) }^{\pi(i) -1} \theta \{ k -
 \phi[ \pi(i) -1 ] ( 1 - \epsilon ) - f [ \pi(i) , \epsilon ] - 1
 \} \delta_{ (
    -1)^{\pi(i)-1+k},1}  \binom{\pi(i)-1}{\frac{\pi(i)-1+k}{2}} \frac{1}{2^{\pi(i)-1}}  + \\
  \frac{1}{2} \sum_{k= - ( \pi(i) -1 ) }^{\pi(i) -1} \theta \{
 \phi[ \pi(i) -1 ] ( 1 + \epsilon ) - k + g [ \pi(i) , \epsilon ] - 1
 \} \delta_{ (
    -1)^{\pi(i)-1+k},1}  \binom{\pi(i)-1}{\frac{\pi(i)-1+k}{2}} \frac{1}{2^{\pi(i)-1}} ] +
    \\
    \theta [ -1 -\pi(i-1) + \pi(i) ] ( 1 - \delta_{-1 - \pi(i-1) +
    \pi(i),0} ) \sum_{k=-  \pi(i)  }^{\pi(i) } \theta [
    k  -  \phi (\pi(i) ) ( 1 -
    \epsilon ) ]   \theta [  \phi (\pi(i) )  ( 1 +
    \epsilon) - k  ]
    \delta_{ (-1)^{\pi(i) +k},1}  \binom{n}{\frac{\pi(i) +k}{2}}
\frac{1}{2^{\pi(i) }} \} \\ \forall m \in \{ 0 , \cdots , N \}
\end{multline}

Such a probability distribution may be implemented algorithmically
through the following Mathematica 5 \cite{Wolfram-03} expressions
(using the new version of the package Combinatorica extensively
illustrated in \cite{Pemmaraju-Skiena-03}):

\begin{verbatim}
Theoretical probability

<<DiscreteMath`Combinatorica`

heaviside[x_]:=If[Element[x,Reals],If[GreaterEqual[x,0],1,0],0]

randomwalkdistribution[n_,k_]:=
  KroneckerDelta[(-1)^(n+k),1] *Binomial[n,(n+k)/2]*2^(-n)

phi[n_]:=Sqrt[2*n*Log[Log[n]]]

f[n_,epsilon_]:=(phi[n]-phi[n-1])*(1-epsilon)

g[n_,epsilon_]:=(phi[n]-phi[n-1])*(1+epsilon)

elementaryprobability[n_,epsilon_]:=
  If[Or[n\[Equal]1,n\[Equal]2],0,
    Sum[ heaviside[k-phi[n] *(1-epsilon) ] *
        heaviside[phi[n] *(1+epsilon)-k ]*
        randomwalkdistribution[n,k]  , {k,-n,n } ]]

consecutiveconditionalprobability[n_,
    epsilon_]:= (1/2)*
      Sum[heaviside[k-phi[n-1]*(1-epsilon)-f[n,epsilon]-1]*
          randomwalkdistribution[n-1,k]                   ,{k,-(n-1),
          n-1 }]+(1/2)*
      Sum[heaviside[phi[n-1]*(1+epsilon)-k+g[n,epsilon]-1]*
          randomwalkdistribution[n-1,k]                   ,{k,-(n-1),n-1 }]

conditionalprobability[n_,m_,epsilon_]:=
  If[m\[Equal]n-1,consecutiveconditionalprobability[n,epsilon],
    elementaryprobability[n,epsilon]]

p[n_,m_]:=KSubsets[Range[3,n],m]

counterdistributionformpositive[n_,epsilon_,m_]:=
  N[Sum[elementaryprobability[p[n,m][[s]][[1]],epsilon]*
        Product[conditionalprobability[p[n,m][[s]][[i]],p[n,m][[s]][[i-1]],
            epsilon],{i,2,m}],{s,1,Length[p[n,m]]}]]

counterdistribution[nbig_,epsilon_,m_]:=
  If[m>0,counterdistributionformpositive[nbig,epsilon,m],
    N[(1-elementaryprobability[3,epsilon])*
        Product[  (1/2)*
              Sum[ heaviside[
                    k-phi[n-1]*(1+epsilon)-g[n,epsilon]-1]*(1-
                      KroneckerDelta[k-phi[n-1]*(1+epsilon)-g[n,epsilon]-1,
                        0])* randomwalkdistribution[n-1,
                    k]          ,{k,-(n-1),n-1} ]+(1/2)*
              Sum[heaviside[
                    phi[n-1]*(1-epsilon)-k+f[n,epsilon]-1]*(1-
                      KroneckerDelta[
                        phi[n-1]*(1-epsilon)-k+f[n,epsilon]-1   ,0])*
                  randomwalkdistribution[n-1,k]       ,{k,-(n-1),n-1}  ],{n,4,
            nbig} ]]]

visualizecounterdistribution[n_,epsilon_]:=
    ListPlot[Table[{i,counterdistribution[n,epsilon,i]},{i,0,n}],
      PlotStyle\[Rule]PointSize[0.02]];
\end{verbatim}

The $ n := 100 $ experiments performed (with $ N := 10^{4} $ and $
\epsilon := 0.1 $) consisting in measuring indirectly the counting
for each of the n binary strings reported in the appendix
\ref{sec:Experimental data} may be elaborated through the
following Mathematica5 expressions:
\begin{verbatim}
Elaboration of the experimental data

epsilon=0.1;

f[x_]:=2*x-1

predicateQ[x_]:=IntervalMemberQ[Interval[
{1-epsilon,1+epsilon}],x];

Do[rescaledstring[n]=Map[f,string[n]];
  iteratedlogarithmlist[n]=
    Table[N[Sum[ Part[rescaledstring[n],i],{i,1,k}]/Sqrt[2*k*Log[Log[k]]]],{k,
        3,10^4}];
  counter[n]=Length[Select[iteratedlogarithmlist[1],predicateQ]], {n,1,100}]
\end{verbatim}

to furnish the following table reporting in the first column the
experimental outcomes of the counter variable $ c_{N,\epsilon} $
and in the right column their absolute frequencies:

\bigskip

\begin{tabular}{|c|c|}
  \hline
  m  & $ f_{m} $ \\ \hline
  0 & 44 \\
  1 & 11 \\
  3 &  2 \\
  4 &  3 \\
  5 &  1 \\
  6 &  3 \\
  7 &  3 \\
  8 &  1 \\
 12 &  1 \\
 15 &  1 \\
 17 &  1 \\
 20 &  1 \\
 23 &  1 \\
 29 &  3 \\
 39 &  1 \\
 40 &  1 \\
 52 &  1 \\
 55 &  1 \\
 61 &  1 \\
 79 &  1 \\
 86 &  2 \\
 128 & 1 \\
 153 & 1 \\
 184 & 1 \\
 207 & 1 \\
 224 & 1 \\
 236 & 1 \\
 259 & 1 \\
 303 & 1 \\
 318 & 1 \\
 335 & 1 \\
 458 & 1 \\
 540 & 1 \\
 1040 & 1 \\
 1044 & 1 \\
 2023 & 1 \\
 2644 & 1 \\ \hline
\end{tabular}

\bigskip

Let us now look at the experimental frequencies $ \{ f_{m}
\}_{m=0}^{N}$  as experimental values of frequency random
variables $  \{ F_{m} \}_{m=0} ^{N}$ clearly having multinomial
joint probability distribution:
\begin{equation}
    P ( \wedge_{m=0}^{N} F_{m} = f_{m} ) \; = \; \delta_{\sum_{m=0}^{N} f_{m},n}  \frac{n ! }{\prod_{m=0}^{N} f_{m}
    !} \prod_{m=0}^{N} P( c_{N,\epsilon} = m)^{f_{m}}
\end{equation}

Let us then introduce the joint characteristic function of the
random variables $  \{ F_{m} \}_{m=0} ^{N}$:
\begin{equation}
    Z( t_{0} , \cdots , t_{N} ) \; := \; E[ e^{i \sum_{m=0}^{N} t_{m}
    f_{m}} ] \; = \; ( \sum_{m=0}^{N} P( c_{N,\epsilon} = m) e^{i
    t_{m}} )^{n}
\end{equation}
and:
\begin{equation}
    W( t_{0} , \cdots , t_{N} ) \; := \; \log  Z( t_{0} , \cdots , t_{N} )
\end{equation}
A simple computation \cite{Cramer-46}, \cite{Shao-03},
\cite{Lehmann-Romano-05} shows that:
\begin{equation} \label{eq:asymptotic expression for the W}
 W( t_{0} , \cdots , t_{N} ) \; = \; e^{- \frac{1}{2} \vec{t}
 \cdot A \vec{t} } + O ( \frac{1}{\sqrt{n}} )
\end{equation}
where I have introduced the compact notation:
\begin{equation}
    \vec{t} \; := ( t_{0} , \cdots , t_{N} )
\end{equation}
\begin{equation}
    \vec{x} \cdot \vec{y} \; := \sum_{m=0}^{N} x_{m} y_{m}
\end{equation}
and where A is the $ ( N+1 ) \times ( N+1) $ matrix:
\begin{equation}
  A \; := \; \mathbb{I} - \sqrt{\vec{p}_{N,\epsilon} } \cdot \sqrt{\vec{p}_{N,\epsilon}}
\end{equation}
with $ \mathbb{I} $ being the $ ( N+1 ) \times ( N+1) $ identity
matrix and $ \sqrt{\vec{p}_{N,\epsilon} } $ being the vector:
\begin{equation}
    \sqrt{\vec{p}_{N,\epsilon} } \; := ( \sqrt{P( c_{N,\epsilon} =
    0)} , \cdots , \sqrt{P( c_{N,\epsilon} = m)} )
\end{equation}

An immediate consequence of the equation \ref{eq:asymptotic
expression for the W} is the Pearson's Theorem stating that, under
the statistical \emph{null hypothesis} $ H_{0} $  that our
experimental data are a sample of values of the random variable $
c_{N,\epsilon} $ with the probability distribution of the equation
\ref{eq:counter distribution}, the random variable:
\begin{equation} \label{eq:chi squared experimental 1}
    \chi^{2}_{exp,1} \; := \; \sum_{m=0}^{N} \frac{[F_{m}- n \cdot P ( c_{N,\epsilon} = m ) ]^{2}}{n \cdot P ( c_{N,\epsilon}= m)}
\end{equation}
tends, in the limit $ n \rightarrow \infty $, to the probability
distribution of a $ \chi^{2}$ random variable with $( N +1) -1 =
N$ degrees of freedom:
\begin{equation}
    \lim_{n \rightarrow + \infty} P( \chi^{2}_{exp} \leq x ) \; =
    \; \int_{- \infty}^{+ x}  \rho_{N}(y) \, dy \; \; \forall x
    \in (0 , + \infty )
\end{equation}
 where a $ \chi^{2}$ random
variable with k degrees of freedom is  a random variable over the
real axis $ \mathbb{R} $ having probability density:
\begin{equation}
    \rho_{k}(x) \; := \; \left\{%
\begin{array}{ll}
    \frac{1}{2^{\frac{k}{2}} \Gamma ( \frac{k}{2} ) } x^{\frac{k}{2}-1} e^{- \frac{x}{2}} , & \hbox{if $ x > 0$;} \\
    0, & \hbox{if $ x \leq 0$.} \\
\end{array}%
\right.
\end{equation}

A rule of thumb generally accepted in Statistics consists in
assuming that, always under the statistical \emph{null hypothesis}
$ H_{0} $  that our experimental data are a sample of values of
the random variable $ c_{N,\epsilon} $ with the probability
distribution of the equation \ref{eq:counter distribution},  the
probability distribution of a $ \chi^{2}$ random variable with N
degrees of freedom is an acceptable approximation of the
probability distribution of $ \chi^{2}_{exp,1} $ provided:
\begin{equation} \label{eq:condition of applicability of the Pearson test}
   n \cdot  P ( c_{N,\epsilon} = m ) \; \geq \; 10 \; \; \forall
   m \in \{ 0 , \cdots , N \}
\end{equation}

Anyway, in our case, an explicit computation shows that:
\begin{equation}
    P ( c_{10^4, 0.1 } = 0 ) \leq P ( C_{100, 0.1} = 0 ) \sim 10^{-33}
\end{equation}

so that, to satisfy the condition \ref{eq:condition of
applicability of the Pearson test}, I should increase the number
of experiments from $ 10^{2} $ to $ 10^{34} $, an obviously
unfeasible task.

Hence in our case the terms  of order $ O ( \frac{1}{\sqrt{n}} ) $
in the equation \ref{eq:asymptotic expression for the W} are not
negligible and must to be taken into account.

Unfortunately the computation of the probabilities appearing in
the right hand side of the equation \ref{eq:chi squared
experimental 1}  through the expressions
\verb"counterdistribution[10^4, 0.1, m]" would take busy the
Mathematica 5 kernel for an enormous amount of time, (as one can
infer from the fact that it involves a sum over the $
\binom{10^4}{m}$  different subsets of m elements of the set $ \{
1 , \cdots 10^4 \} $) and is therefore practically unfeasible.

 A way to bypass the problem consists in the introduction of:
\begin{enumerate}
    \item the following two events:
 \begin{equation}
    S_{1} \; := \; c_{10^4,0.1} = 0
\end{equation}
\begin{equation}
    S_{2} \; := \; c_{10^4,0.1} \neq  0
\end{equation}
\item their theoretical probabilities:
\begin{equation}
   p_{1} \; := \; p \; := P(S_{1})
\end{equation}
\begin{equation}
   p_{2} \; := \; 1-p \; := P(S_{2})
\end{equation}
    \item their  experimental frequencies:
\begin{equation}
    \tilde{f}_{1,exp} \; := \; \tilde{f}_{exp} \; :=  \sum_{m : S_{1} \text{ holds} } f_{m} \; = \;
    44
\end{equation}
\begin{equation}
    \tilde{f}_{2,exp} \; := \; ( n - \tilde{f}_{exp} ) \; := \; \sum_{m : S_{2} \text{ holds} } f_{m} \; = \;
    56
\end{equation}
\end{enumerate}

Let us now look at the experimental frequencies $\{
\tilde{f}_{exp} , n - \tilde{f}_{exp} \} $ as experimental values
of frequency random variables $\{ \tilde{F} ,  n - \tilde{F} \} $
that, under the statistical \emph{null hypothesis} $ H_{0} $  that
our experimental data are a sample of values of the random
variable $ c_{N,\epsilon} $ with the probability distribution of
the equation \ref{eq:counter distribution}, have the following
binomial probability distribution:
\begin{equation}
    P ( \tilde{F} = \tilde{f})  \; = \;  \binom{n}{ \tilde{f} } p^{\tilde{f}} ( 1 - p)^{n - \tilde{f}}
\end{equation}

Introduced the random variable:
\begin{equation}
    \chi^{2}_{exp,2} \; := \; \frac{ ( \tilde{F} - n p )^{2}  }{ n p
    } \, + \, \frac{ [ (n- \tilde{F})  - n ( 1 - p )  ]^{2}  }{ n ( 1
    - p )} \; = \; \frac{ (\tilde{F} - n p)^{2}}{n p ( 1 - p) }
\end{equation}
taking values in the set $ \{  \frac{( \tilde{f} - n p )^{2}  }{ n
p} \, , \,  \tilde{f} \in \{ 0 , \cdots , n \} \}  $  let us
observe that:
\begin{equation} \label{eq:cumulative probability distribution of the experimental chi squared2}
    P (   \chi^{2}_{exp,2}  \leq x ) \; = \sum_{\tilde{f} = 0}^{n}
    \theta [ x -  \frac{ (\tilde{f} - n p)^{2}}{n p ( 1 - p) } ]  P ( \tilde{F} = \tilde{f})
     \; = \\
  \sum_{\tilde{f} = 0}^{n} \theta [ x -  \frac{ (\tilde{f} - n p)^{2}}{n p ( 1 - p) } ]
\binom{n}{ \tilde{f} } p^{\tilde{f}} ( 1 - p)^{n - \tilde{f}}
\end{equation}

Choosen a \emph{confidence coefficient} $ \alpha  $ let us
introduce the \emph{quantile of order $ \alpha $} of the
probability distribution $ \chi^{2}_{exp,2} $:
\begin{equation}
    q_{\alpha} \; := \; \min \{ \tilde{f} \in \{ 0 , \cdots , n \} : P (   \chi^{2}_{exp,2}  \leq \tilde{f} ) \geq \alpha
    \}
\end{equation}

Then:
\begin{enumerate}
    \item if  $ \chi^{2}_{exp,2} ( \tilde{f}_{exp} ) \leq
    q_{\alpha}$ then the \emph{null hypothesis} $ H_{0} $ has to be
    accepted with \emph{confidence coefficient} $ \alpha  $
    \item if $ \chi^{2}_{exp,2} ( \tilde{f}_{exp} ) >
    q_{\alpha} $ hen the \emph{null hypothesis} $ H_{0} $ has to be
    rejected with \emph{confidence coefficient} $ \alpha  $
\end{enumerate}

By the equation \ref{eq:cumulative probability distribution of the
experimental chi squared2} we obtain that:
\begin{equation}
   q_{\alpha} \; = \;   \max \{ \tilde{f} \in \{ 0 , \cdots , n \} :  \sum_{\tilde{f} = 0}^{n} \theta [ \alpha -  \frac{ (\tilde{f} - n p)^{2}}{n p ( 1 - p) } ]
\binom{n}{ \tilde{f} } p^{\tilde{f}} ( 1 - p)^{n - \tilde{f}} \leq
\alpha  \}
\end{equation}

Fixed the \emph{confidence coefficient} $ \alpha:= 1 - 10^{ - 10 }
$ one can easily use the following Mathematica 5 expressions:

\begin{verbatim}
Statistical test

alpha=1-10^(-10);

binomialdistribution[n_,p_,k_]:=Binomial[n,k]*(p^k)*((1-p)^(n-k))

chisquared[n_,p_,f_]:=(f-n*p)^2/(n*p*(1-p))

cumulativechisquared[n_,p_,x_]:=
  N[Sum [UnitStep[ x-chisquared[n,p,f]]*binomialdistribution[n,p,f] ,{f,0,n}],
    1000]

predicateQ[x_]:=GreaterEqual[x[[2]],alpha]

sightseeing[n_,p_]:=Table[ {f,cumulativechisquared[n,p,f]}
,{f,0,n}]

myquantile[n_,p_]:=Select[sightseeing[n,p],predicateQ][[1]][[1]]

test[n_,p_,f_]:=LessEqual[chisquared[n,p,f],myquantile[n,p]]
\end{verbatim}
to observe that, since $ p \leq 10^{-33} < \frac{11}{25} $, it
follows that $ \chi^{2}_{exp,2} ( \tilde{f}_{exp} ) \geq 10^{34} $
and hence it is enormously greater than $ q_{1-10^{-10}} = 1 $;
therefore we can reject the null hypothesis $ H_{0} $ with
confidence coefficient $ 1 - 10^{-10} $.

So, under the assumption that \emph{Quantis} is a quantum coin, it
follows that the quantum coin doesn't pass the Iterated Logarithm
Randomness test with confidence coefficient $ 1 - 10^{-10} $.

\newpage
Let us now remark that, owing to the proposition
\ref{prop:algorithmic foundation of Probability Theory}, the Axiom
of Reduction (i.e. the axiom \ref{ax:Axiom of Reduction of Quantum
Mechanics}) implies the following:

\begin{axiom} \label{ax:Martin Lof - Solovay - Chaitin Axiom of Reduction}
\end{axiom}
\emph{Martin L\"{o}f - Solovay - Chaitin Axiom of Reduction:}

\begin{hypothesis}
\end{hypothesis}
\begin{center}
  Let $ x_{n} $ be the result of the measurement of the qubit
 operator on the $ n^{th}$ quantum coin (prepared in the state $ |
 unbiased > $).
\end{center}
\begin{thesis}
\end{thesis}
\begin{enumerate}
    \item
\begin{equation}
    \{ x_{n} \}_{n \in \mathbb{N}_{+}} \in  RANDOM_{MLSC} ( \{ -1 , 1
    \}^{\mathbb{N}_{+}})
\end{equation}
    \item if the result of the measurement of $  \hat{q} $ on the $ n^{th} $ quantum coin is
    equal to -1 then the state of the  $ n^{th} $  quantum coin under the
    measurement is $ |0 > $; if, contrary, the result of the measurement of $  \hat{q} $ on the $ n^{th} $ quantum coin  is
    equal to -1 then the state of the $ n^{th} $ quantum coin under the
    measurement is $ |1 > $.
\end{enumerate}

Owing to the proposition \ref{prop:Martin Lof-Solovay Chaitin
random sequences obey the Law of the Iterated Logarithm} I have
shown the experimental violation of the axiom \ref{ax:Martin Lof -
Solovay - Chaitin Axiom of Reduction} and, consequentially, of the
axiom \ref{ax:Axiom of Reduction of Quantum Mechanics}.

\newpage
\section{The Von Mises - Church Axiom of Reduction}
 The experimental indications of the violation of the Martin L\"{o}f - Solovay - Chaitin Axiom of Reduction (i.e. the axiom \ref{ax:Martin Lof - Solovay - Chaitin Axiom of
 Reduction}) shown in the section \ref{sec:Experimental violation of the Iterated Logarithm
randomness test} naturally suggest to replace such an axiom with
the
 following weaker:

\begin{axiom} \label{ax:Von Mises-Church Axiom of Reduction}
\end{axiom}
\emph{Von Mises - Church Axiom of Reduction:}

\begin{hypothesis}
\end{hypothesis}
\begin{center}
  Let $ x_{n} $ be the result of the measurement of the qubit
 operator on the $ n^{th}$ quantum coin (prepared in the state $ |
 unbiased > $).
\end{center}
\begin{thesis}
\end{thesis}
\begin{enumerate}
    \item
\begin{equation}
    \{ x_{n} \}_{n \in \mathbb{N}_{+}} \in  RANDOM_{VMC} ( \{ -1 , 1
    \}^{\mathbb{N}_{+}})
\end{equation}
    \item if the result of the measurement of $  \hat{q} $ on the $ n^{th} $ quantum coin is
    equal to -1 then the state of the  $ n^{th} $  quantum coin after the
    measurement is $ |0 > $; if, contrary, the result of the measurement of $  \hat{q} $ on the $ n^{th} $ quantum coin  is
    equal to 1 then the state of the $ n^{th} $ quantum coin after the
    measurement is $ |1 > $.
\end{enumerate}

Let us remark that the replacement of the Martin L\"{o}f - Solovay
- Chaitin Axiom of Reduction (i.e. the axiom \ref{ax:Martin Lof -
Solovay - Chaitin Axiom of
 Reduction}) with the Von - Mises Church Axiom of Reduction (i.e. the axiom \ref{ax:Von Mises-Church Axiom of Reduction}) corresponds, ironically, to a return to the viewpoint of  the founding fathers
of Quantum Mechanics who,  writing in days in which the sound
Kolmogorovian foundation of (Classical) Probability Theory ,
contained in the book \cite{Kolmogorov-56} first published in
1933, was still lacking, actually assumed a version of the
Postulate of Reduction someway oscillating between something
similar to the Von - Mises Church Axiom of Reduction (i.e. a Von
Mises' like approach without a precise definition of the
"\emph{admissible place selection rules}"; see for instance the
celebrated Von Neumann's book \cite{Von-Neumann-83} first
published in 1932) and the following radically minimal version
(see for instance Dirac's celebrated book \cite{Dirac-58} first
published in 1930):
\begin{axiom} \label{ax:minimal axiom of reduction}
\end{axiom}
\emph{Minimal Axiom of Reduction:}

\begin{hypothesis}
\end{hypothesis}
\begin{center}
  Let $ x_{n} $ be the result of the measurement of the qubit
 operator on the $ n^{th}$ quantum coin (prepared in the state $ |
 unbiased > $).
\end{center}
\begin{thesis}
\end{thesis}
\begin{enumerate}
    \item
\begin{equation}
    \{ x_{n} \}_{n \in \mathbb{N}_{+}} \text{ is Borel-1-normal }
\end{equation}
    \item if the result of the measurement of $  \hat{q} $ on the $ n^{th} $ quantum coin is
    equal to -1 then the state of the  $ n^{th} $  quantum coin after the
    measurement is $ |0 > $; if, contrary, the result of the measurement of $  \hat{q} $ on the $ n^{th} $ quantum coin  is
    equal to 1 then the state of the $ n^{th} $ quantum coin after the
    measurement is $ |1 > $.
\end{enumerate}
where we recall that given an alphabet A, a sequence $ \bar{x} \in
A^{\mathbb{N}_{+}} $ and a positive integer $  m \in
\mathbb{N}_{+} $:
\begin{definition}
\end{definition}
\emph{$ \bar{x} $ is Borel m-normal:}
\begin{equation}
    \lim_{n \rightarrow \infty} \frac{N_{a}^{m} ( \vec{x}(n))}{ \lfloor \frac{n}{m} \rfloor
    } \; = \; \frac{1}{ |A|^{m} } \; \; \forall a \in A
\end{equation}
where $ \lfloor x \rfloor \; := \; \max \{ n \in \mathbb{Z} : n
\leq x \} $ is the \emph{floor} of the real number x.
\begin{definition}
\end{definition}
\emph{$ \bar{x} $ is Borel normal:}
\begin{equation}
  \bar{x} \text{ is Borel-m-normal } \; \forall m \in \mathbb{N}_{+}
\end{equation}

Since obviously:
\begin{proposition} \label{prop:Von Mises-Church randomness implies Borel 1-normality}
\end{proposition}
\emph{Von Mises - Church randomness implies Borel 1-normality:}
\begin{equation}
    \bar{x} \text{ is Borel 1-normal } \; \; \forall \bar{x} \in
    RANDOM_{VMC}( A^{\mathbb{N}_{+}} )
\end{equation}

the Minimal Axiom of Reduction (i.e. the axiom \ref{ax:minimal
axiom of reduction}) is weaker than the Von Mises - Church Axiom
of Reduction (i.e. the axiom \ref{ax:Von Mises-Church Axiom of
Reduction})

 Actually most all the impressive experimental
confirmations of Quantum Mechanics, being based only on the
measurements of the relative frequencies of experimental outcomes,
are confirmations only of the  Minimal Axiom of Reduction.

Such an axiom is, anyway, too weak since the fact that independent
tosses of the quantum coin pass a huge amount of randomness
statistical tests (such as the absence of correlations observed by
Berkeland, Raymondson and Tassin
\cite{Berkeland-Raymondson-Tassin-04} or the fact that, according
to the test report contained in the enclosed material, the quantum
random number generator \emph{Quantis} passes both the NIST test
\cite{Nist-08} and the DIEHARD test \cite{Diehard-08}) may be seen
as confirmations of both  the stronger Von Mises - Church Axiom of
Reduction and  Martin Lof - Solovay - Chaitin Axiom of Reduction
(though they are not able to detect the difference betweeen these
two axioms).

In order to appreciate the game theoretic meaning underlying the
Von Mises Church Axiom of Reduction let us denote by $
A_{Alice}^{quantum} \subset A_{Alice} $ the set of the Alice's
actions in the Von Mises - Church game obtained by a countable
infinity of independent tosses of the quantum coin.

Then:

\begin{proposition} \label{prop:quantum coin versus Nash equilibrium}
\end{proposition}
\emph{Quantum randomness versus Nash equilibrium:}

\begin{hypothesis}
\end{hypothesis}

\begin{center}
  the axiom \ref{ax:Von Mises-Church Axiom of Reduction} is assumed
\end{center}

\begin{thesis}
\end{thesis}
\begin{equation}
  ( A_{Alice}^{quantum} \times A_{Bob} ) \cap \mathcal{E}_{Nash}(
  \mathcal{G}_{VMC} ) \; \neq \; \emptyset
\end{equation}

\begin{proof}
The thesis immediately follows  by the proposition \ref{prop:Von
Mises Church randomness versus Nash equilibrium}.
\end{proof}

\newpage
\appendix
\section{Experimental data} \label{sec:Experimental data}
After having generated $10^2$ binary strings generated by
\emph{Quantis}, each of length $10^4 $, I have acquired them
through a  Mathematica 5 \cite{Wolfram-03} session in which the $
n^{th} $ string is encoded into the variable \verb"string[n]".

I report, to save space, their conversion in base 10:



\newpage
\section{Notation}

 \bigskip
\begin{tabular}{|c|c|}
  \hline
  i.e. & id est \\
  HP: & hypothesis \\
  TH: & thesis \\
  $ \forall $ & for all (universal quantificator) \\
  $ \exists $ & exists (existential quantificator) \\
  $ \exists \; ! $ & exists and is unique \\
  $ x \; = \; y $ & x is equal to y \\
  $ x \; := \; y $ & x is defined as y \\
\hline
 \end{tabular}

\end{document}